\documentclass[conference,10pt]{IEEEtran}
\IEEEoverridecommandlockouts

\usepackage{epsfig,rotating,setspace,latexsym,amsmath,epsf,amssymb,amsfonts,bm,theorem,cite,algorithm,graphicx,epsf,authblk,epstopdf,color,algpseudocode,bbm}

\newtheorem{theorem}{Theorem}
\newtheorem{lemma}{Lemma}

\allowdisplaybreaks

\begin{document}

\title{Near Optimal Online Distortion Minimization\\for Energy Harvesting Nodes\thanks{This work was supported by NSF Grants CNS 13-14733, CCF 14-22111, CCF 14-22129, and CNS 15-26608.}}

\author{Ahmed Arafa \quad Sennur Ulukus\\
\normalsize Department of Electrical and Computer Engineering\\
\normalsize University of Maryland, College Park, MD 20742\\
\normalsize \emph{arafa@umd.edu} \quad \emph{ulukus@umd.edu}}

\maketitle

\begin{abstract}
We consider online scheduling for an energy harvesting communication system where a sensor node collects samples from a Gaussian source and sends them to a destination node over a Gaussian channel. The sensor is equipped with a finite-sized battery that is recharged by an independent and identically distributed (i.i.d.) energy harvesting process over time. The goal is to minimize the long term average distortion of the source samples received at the destination.  We study two problems: the first is when sampling is cost-free, and the second is when there is a sampling cost incurred whenever samples are collected. We show that {\it fixed fraction policies} \cite{ozgur_online_su}, in which a fixed fraction of the battery state is consumed in each time slot, are near-optimal in the sense that they achieve a long term average distortion that lies within a constant {\it additive} gap from the optimal solution for all energy arrivals and battery sizes. For the problem with sampling costs, the transmission policy is bursty; the sensor can collect samples and transmit for only a portion of the time.
\end{abstract}

\section{Introduction}

A sensor node collects samples from an i.i.d. Gaussian source and sends them to a destination over a Gaussian channel. The sensor relies solely on energy harvested from nature and is equipped with a finite-sized battery to save its incoming energy. The goal is to characterize {\it online} power control policies that minimize the long term average distortion of the received samples at the destination.

{\it Offline} power scheduling in energy harvesting communication systems has been extensively studied in the recent literature. Earlier works \cite{jingP2P, kayaEmax, omurFade, ruiZhangEH} consider the single-user setting. References \cite{jingBC, elifBC, omurBC, jingMAC, aggarwalPmax, onurCoopEH, kaya-interference} extend this to broadcast, multiple access, and interference settings; and \cite{ruiZhangRelay, gunduz2hop, letaiefRelay, berkDiamond-jour} consider two-hop and relay channels. Energy sharing and energy cooperation concepts are studied in \cite{berkCoop, kayaProcrastinate}. References \cite{kayaRxEH, yatesRxEH1, payaroRxEH, rahulOnOff, arafaJSACdec, arafa_baknina_twc_dec_proc} study energy harvesting receivers, where energy harvested at the receiver is spent mainly for sampling and decoding. Other works \cite{ruiZhangNonIdeal, orhan-broadband, omurHybrid, payaro-cost, baknina_temp, arafa_baknina_twc_proc} study the impact of processing costs, i.e., the power spent for circuitry, on energy harvesting communications. A source-channel coding problem with an energy harvesting transmitter is formulated in \cite{orhan_dist} to minimize the distortion of source samples sent to a destination. Impacts of processing and sampling costs are also studied, and two-dimensional water-filling interpretations are presented.

Recently, \cite{ozgur_online_su} has introduced an online power control policy for a single-user energy harvesting channel that maximizes the long term average throughput. The proposed policy is near optimal in the sense that it performs within a constant gap from the optimal solution that is independent of energy arrivals and battery sizes. This is extended to broadcast channels in \cite{baknina_online_bc}, multiple access channels in \cite{ozgur_online_mac, baknina_online_mac}, and systems with processing costs in \cite{baknina_online_su_proc, baknina_online_twc_proc}.

In this paper, we follow the approaches in \cite{ozgur_online_su, baknina_online_bc, ozgur_online_mac, baknina_online_mac, baknina_online_su_proc, baknina_online_twc_proc} to extend the offline results in \cite{orhan_dist} to online settings. We characterize near-optimal online power control policies that minimize the long term average distortion in a single-user energy harvesting channel, where a sensor node collects samples from an i.i.d. Gaussian source and sends them to a destination over a Gaussian channel. The sensor relies on energy harvested from nature to transmit its packets and is equipped with a finite-sized battery. Energy is harvested in packets following an i.i.d. distribution with amounts known causally at the sensor, and is consumed in sampling and transmission. We formulate two problems: one with and the other without sampling energy consumption costs. In both problems, we show that the policy introduced in \cite{ozgur_online_su} achieves a long term average distortion that lies within a constant additive gap from the optimal achieved distortion for all energy arrivals and battery sizes.

\section{System Model and Problem Formulation}

We consider a sensor node collecting i.i.d. Gaussian source samples, with zero-mean and variance $\sigma_s^2$, over a sequence of time slots. Without loss of generality, a slot duration is normalized to one time unit. Samples are compressed and sent over an additive white Gaussian noise channel, with variance $\sigma_c^2$, to an intended destination. We consider a strict delay scenario where samples need to be sent during the same time slot in which they are collected. With a mean squared error distortion criterion, the average distortion of the source samples in time slot $t$, $D_t$, is given by \cite{cover}
\begin{align} \label{eq_d_t}
D_t=\sigma_s^2\exp{(-2r_t)}
\end{align}
where $r_t$ denotes the sampling rate at time slot $t$. 

The sensor uses energy harvested from nature to send its samples over the channel, with minimal distortion. Energy arrives (is harvested) in packets of amount $E_t$ at the beginning of time slot $t$, and follows an i.i.d. distribution with a given mean. Our setting is {\it online,} in the sense that the amounts of energy are known causally in time, i.e., after being harvested. Only the mean of the energy arrivals is known a priori. Energy is saved in a battery of finite size $B$. The sensor consumes energy in sampling and transmission. Depending on the physical settings, sampling energy can be a significant system aspect and needs to be taken into consideration \cite{orhan_dist}. We formulate two different problems for that matter: one without, and the other with sampling costs as follows.

\subsection{No Sampling Costs Case}

Let $\mathcal{E}^t\triangleq\{E_1,E_2,\dots,E_t\}$, and let $g_t$ denote the the sensor's transmission power in time slot $t$. A feasible online policy ${\bm g}$ is a sequence of mappings $\left\{g_t:~\mathcal{E}^t\rightarrow \mathbb{R}_+\right\}$ satisfying
\begin{align}
g_t\leq b_t\triangleq\min\{b_{t-1}-g_{t-1}+E_t,B\},\quad \forall t
\end{align}
with $b_1\triangleq B$ without loss of generality (using similar arguments as in \cite[Appendix B]{ozgur_online_su}). We denote the feasible set above by $\mathcal{F}$. By allocating power $g_t$ at time slot $t$ to the Gaussian channel, the sensor achieves an instantaneous communication rate of \cite{cover}
\begin{align} \label{eq_r_t}
r_t=\frac{1}{2}\log\left(1+g_t/\sigma_c^2\right)
\end{align}
Given a feasible policy ${\bm g}$, and using (\ref{eq_d_t}) and (\ref{eq_r_t}), we define the $n$-horizon average distortion as
\begin{align}
\mathcal{D}_n({\bm g})\triangleq\frac{1}{n}\mathbb{E}\left[\sum_{t=1}^n\frac{\sigma_s^2}{1+g_t/\sigma_c^2}\right]
\end{align}
Our goal is to minimize the long term average distortion, subject to (online) energy causality constraints. That is, to characterize the following
\begin{align} \label{opt_dist}
d^*\triangleq\min_{{\bm g}\in\mathcal{F}}\lim_{n\rightarrow\infty}\mathcal{D}_n({\bm g})
\end{align}

\subsection{The Case With Sampling Costs}

Next, we consider the case where sampling the source incurs an energy cost $\epsilon$ per unit time, that is a constant independent of the sampling rate. Due to the sampling cost, collecting all the source samples might not be optimal. Hence, we allow the sensor to be {\it on} during a $\theta_t\leq1$ portion of time slot $t$, and turn off for the remainder of the time slot. The expected distortion achieved in time slot $t$ under this setting is now given by
\begin{align} \label{eq_d_t_eps}
D_t^\epsilon=(1-\theta_t)\sigma_s^2+\theta_t\sigma_s^2\exp{(-2r_t)}
\end{align}
and the feasible set $\mathcal{F}^\epsilon$ is now given by the sequence of mappings $\{(\theta_t,g_t):~\mathcal{E}^t\rightarrow [0,1]\times\mathbb{R}_+\}$ satisfying
\begin{align}
\theta_t(\epsilon+g_t)\leq b_t\triangleq\min\{b_{t-1}-\theta_{t-1}(\epsilon+g_{t-1})+E_t,B\},~\forall t
\end{align}
with $b_1\triangleq B$. We note that the problem with sampling costs is formulated slightly different in \cite{orhan_dist}. In our formulation, the expected distortion is interpreted by time sharing between not transmitting (and hence achieving $\sigma_s^2$) and transmitting with rate $r_t$ (and hence achieving $\sigma_s^2\exp(-2r_t)$). Given a feasible policy $\left({\bm \theta},{\bm g}\right)$, and using (\ref{eq_r_t}) and (\ref{eq_d_t_eps}), we define the $n$-horizon average distortion with sampling costs as
\begin{align}
\mathcal{D}_n^\epsilon\left({\bm \theta},{\bm g}\right)\triangleq\frac{1}{n}\mathbb{E}\left[\sum_{t=1}^n(1-\theta_t)\sigma_s^2+\frac{\theta_t\sigma_s^2}{1+g_t/\sigma_c^2}\right]
\end{align}
whence our goal is to characterize
\begin{align} \label{opt_dist_proc}
d^*_\epsilon\triangleq\min_{({\bm \theta},{\bm g})\in\mathcal{F^\epsilon}}\lim_{n\rightarrow\infty}\mathcal{D}_n^\epsilon\left({\bm \theta},{\bm g}\right)
\end{align}

\section{Main Results}

In this section we discuss the main results of this paper regarding problems (\ref{opt_dist}) and (\ref{opt_dist_proc}). We first note that both problems can be optimally solved via dynamic programming techniques, since the underling system evolves as a Markov decision process. However, the optimal solution is usually computationally demanding with few structural insights. Therefore, in the sequel, we aim at finding relatively simple online power control policies that are provably within a constant additive gap of the optimal solution. Towards that end, we derive lower and upper bounds on the optimal solution and bound the additive gap between them. We show that the proposed policy lies in between these bounds and is therefore of at most the same additive gap from the optimal solution. 

We assume that $E_t\leq B~\forall t$ a.s., since any excess energy above the battery capacity cannot be saved or used. Let $\mu=\mathbb{E}[E_t]$, where $\mathbb{E}[\cdot]$ is the expectation operator, and define
\begin{align} \label{eq_fraction_q}
q\triangleq\frac{\mathbb{E}[E_t]}{B}
\end{align}
Then, we have $0\leq q\leq1$. For problem (\ref{opt_dist}), we define the power control policy as follows \cite{ozgur_online_su}
\begin{align} \label{eq_ffp_dist}
\tilde{g}_t=qb_t
\end{align}
and for problem (\ref{opt_dist_proc}), we define it as
\begin{align} \label{eq_ffp_dist_proc}
\tilde{\theta}_t(\epsilon+\tilde{g}_t)=qb_t 
\end{align}
That is, for either problem, in each time slot, the sensor uses a fixed fraction of its available energy in the battery. Such policies were first introduced in \cite{ozgur_online_su}, and coined {\it fixed fraction policies} (FFP). Clearly these policies are always feasible since $q\leq1$. We note that using (\ref{eq_ffp_dist_proc}) in problem (\ref{opt_dist_proc}) decouples the problem into multiple single-slot problems where the energy consumption in time slot $t$ is $qb_t$. In upcoming sections, we show that solving that single-slot problem for $\left(\tilde{\theta}_t,\tilde{g}_t\right)$ gives
\begin{align} \label{eq_ffp_dist_proc_th}
\tilde{\theta}_t=\min\left\{\frac{qb_t}{\epsilon+\sqrt{\epsilon\sigma_c^2}},1\right\}
\end{align}
\begin{align} \label{eq_ffp_dist_proc_pwr}
\tilde{g}_t=\max\left\{qb_t-\epsilon,\sqrt{\epsilon\sigma_c^2}\right\}
\end{align}
Observe that in the above assignment, for a single energy arrival, either the transmission power or the {\it on} time decreases over slots in a fractional manner, i.e., while one decreases the other one is fixed. Let $d\left(\tilde{{\bm g}}\right)$ and $d_\epsilon\left(\tilde{{\bm \theta}},\tilde{{\bm g}}\right)$ denote the long term average distortion under $\{\tilde{g}_t\}$ in (\ref{eq_ffp_dist}) and $\{(\tilde{\theta}_t,\tilde{g}_t)\}$ in (\ref{eq_ffp_dist_proc_th}) and (\ref{eq_ffp_dist_proc_pwr}), respectively. We now state the main results.

\begin{theorem} \label{thm_dist}
For all i.i.d. energy arrivals with mean $\mu$, the optimal solution of problem (\ref{opt_dist}) satisfies
\begin{align} \label{eq_dist_lower_bound}
d^*\geq f(\mu)\triangleq\frac{\sigma_s^2}{1+\mu/\sigma_c^2}
\end{align}
and the FFP in (\ref{eq_ffp_dist}) satisfies
\begin{align}
f(\mu)\leq d\left(\tilde{{\bm g}}\right)\leq f(\mu)+\frac{1}{2}\sigma_s^2
\end{align}
for all values of $\mu$ and $\sigma_c^2$.
\end{theorem}

\begin{theorem} \label{thm_proc}
For all i.i.d. energy arrivals with mean $\mu$, the optimal solution of problem (\ref{opt_dist_proc}) satisfies
\begin{align} \label{eq_dist_proc_lower_bound}
d_\epsilon^*\geq f_\epsilon(\mu)\triangleq\min_{\theta,\bar{g}} \quad &(1-\theta)\sigma_s^2+\theta\frac{\sigma_s^2}{1+\frac{\bar{g}}{\theta\sigma_c^2}} \nonumber \\
\emph{s.t.} \quad &\theta\epsilon+\bar{g}\leq\mu,\quad0\leq\theta\leq1
\end{align}
and the FFP in (\ref{eq_ffp_dist_proc_th}) and (\ref{eq_ffp_dist_proc_pwr}) satisfies
\begin{align}
f_\epsilon(\mu)\leq d_\epsilon\left(\tilde{{\bm \theta}},\tilde{{\bm g}}\right)\leq f_\epsilon(\mu)+\frac{1}{2}\sigma_s^2
\end{align}
for all values of $\epsilon$, $\mu$, and $\sigma_c^2$.
\end{theorem}

Note that the results in the theorems directly imply that the average long term distortion under the FFP proposed for both problems (\ref{opt_dist_proc}) and (\ref{opt_dist}) lies within a constant additive gap from the optimal solution. We also observe that the additive gap indicated in Theorem~\ref{thm_proc} does not depend on the cost $\epsilon$. We prove these two theorems in the upcoming sections.

\section{No Sampling Costs: Proof of Theorem~\ref{thm_dist}}

\subsection{Lower Bounding $d^*$} \label{sec_lb_dist}

In this section, we derive the lower bound in (\ref{eq_dist_lower_bound}). Following \cite{ozgur_online_su} and \cite{baknina_online_su_proc}, we first remove the battery capacity constraint setting $B=\infty$. This way, the feasible set $\mathcal{F}$ becomes
\begin{align} \label{eq_feas_off}
\sum_{t=1}^ng_t\leq\sum_{t=1}^nE_t,\quad\forall n
\end{align}
Then, we remove the expectation and consider the offline setting of problem (\ref{opt_dist}), i.e., when energy arrivals are known a priori. Since the energy arrivals are i.i.d., the strong law of large numbers indicates that $\lim_{n\rightarrow\infty}\frac{1}{n}\sum_{t=1}^nE_t=\mu$ a.s., i.e., for every $\delta>0$, there exists $n$ large enough such that $\frac{1}{n}\sum_{t=1}^nE_t\leq\mu+\delta$ a.s., which implies by (\ref{eq_feas_off}) that the feasible set, for such $(\delta,n)$ pair, is given by
\begin{align}
\frac{1}{n}\sum_{t=1}^ng_t\leq\mu+\delta\quad\text{a.s.}
\end{align}

Now fix such $(\delta,n)$ pair. The objective function is given by
\begin{align}
\frac{1}{n}\sum_{t=1}^n\frac{\sigma_s^2}{1+g_t/\sigma_c^2}=\frac{1}{n}\sum_{t=1}^nf(g_t)
\end{align}
It is direct to see that $f$ is convex. Therefore, the optimal power allocation minimizing the objective function is $g_t=\mu+\delta$, $1\leq t\leq n$ \cite{boyd} (see also \cite{jingP2P}). Whence, the optimal offline solution is given by $f(\mu+\delta)$. We then have $d^*\geq f(\mu+\delta)$. Since this is true $\forall\delta>0$, we can take $\delta$ down to 0 by taking $n$ infinitely large. Therefore, (\ref{eq_dist_lower_bound}) holds.

\subsection{Upper Bounding $d^*$: Bernoulli Energy Arrivals} \label{sec_ub_dist}

In this section, we derive an upper on $d^*$. Towards that, we first the study a special energy harvesting i.i.d. process: the Bernoulli process. Let $\{\hat{E}_t\}$ be a Bernoulli energy arrival process with mean $\mu$ as follows
\begin{align} \label{eq_bern_energy}
\hat{E}_t\in\{0,B\},~\text{with}~\mathbb{P}[\hat{E}_t=B]=p,~\text{and}~pB=\mu
\end{align}
where $\mathbb{P}[A]$ denotes the probability of $A$. Note that under such specific energy arrival setting, whenever an energy packet arrives, it completely fills the battery, and resets the system. This constitutes a {\it renewal}. Then, by \cite[Theorem 3.6.1]{ross_stochastic} (see also \cite{ozgur_online_su}), the following holds for any power control policy ${\bm g}$
\begin{align} \label{eq_renewal}
\lim_{n\rightarrow\infty}\hat{\mathcal{D}}_n({\bm g})&=\lim_{n\rightarrow\infty}\frac{1}{n}\mathbb{E}\left[\sum_{t=1}^n\frac{\sigma_s^2}{1+g_t/\sigma_c^2}\right] \nonumber \\
&=\frac{1}{\mathbb{E}[L]}\mathbb{E}\left[\sum_{t=1}^L\frac{\sigma_s^2}{1+g_t/\sigma_c^2}\right]\quad\text{a.s.}
\end{align}
where $\hat{\mathcal{D}}_n({\bm g})$ is the $n$-horizon average distortion under Bernoulli arrivals, and $L$ is a random variable denoting the inter-arrival time between energy arrivals, which is geometric with parameter $p$, and $\mathbb{E}[L]=1/p$.

Now, substituting by the FFP (\ref{eq_ffp_dist}) gives an upper bound on $d^*$. Note that by (\ref{eq_bern_energy}), the fraction $q$ in (\ref{eq_fraction_q}) is now equal to $p$. Also note that in between energy arrivals, the battery state decays exponentially, and the FFP in (\ref{eq_ffp_dist}) gives
\begin{align} \label{eq_ffp_bern}
\tilde{g}_t&=p(1-p)^{t-1}B=(1-p)^{t-1}\mu
\end{align}
for all time slots $t$, where the second equality follows since $pB=\mu$. Therefore, using (\ref{eq_renewal}) and (\ref{eq_ffp_bern}), we bound the distortion under the FFP as follows
\begin{align}
\lim_{n\rightarrow\infty}&\hat{\mathcal{D}}_n(\tilde{\bm g})\nonumber \\
&=\frac{1}{\mathbb{E}[L]}\mathbb{E}\left[\sum_{t=1}^L\frac{\sigma_s^2}{1+(1-p)^{t-1}\mu/\sigma_c^2}\right] \nonumber \\
&\stackrel{(a)}{\leq} \frac{1}{\mathbb{E}[L]}\mathbb{E}\left[\sum_{t=1}^L\frac{\sigma_s^2}{1+\mu/\sigma_c^2}+\left(1-(1-p)^{t-1}\right)\sigma_s^2\right] \nonumber \\
&=f(\mu)+\sigma_s^2\left(1-\frac{1}{\mathbb{E}[L]}\mathbb{E}\left[\sum_{t=1}^L(1-p)^{t-1}\right]\right) \nonumber \\
&=f(\mu)+\sigma_s^2\frac{p(1-p)}{1-(1-p)^2} \stackrel{(b)}{\leq} f(\mu)+\frac{\sigma_s^2}{2} \label{eq_ub_dist}
\end{align}
where $(a)$ follows since $\frac{1}{1+\lambda x}\leq\frac{1}{1+x}+(1-\lambda)$ for $0\leq\lambda\leq1$ and $x\geq0$; and $(b)$ follows since $\frac{p(1-p)}{1-(1-p)^2}$ has a maximum value of $1/2$ for $0\leq p\leq1$. Next, we use the above result for Bernoulli arrivals to bound the distortion for general i.i.d. arrivals under the FFP in the following lemma; the proof follows by convexity and monotonicity of $f$, along the same lines of \cite[Section VII-C]{ozgur_online_su}, and is omitted for brevity.

\begin{lemma} \label{thm_bern_iid}
Let $\{\hat{E}_t\}$ be a Bernoulli energy arrival process as in (\ref{eq_bern_energy}) with parameter $q$ as in (\ref{eq_fraction_q}) and mean $qB=\mu$. Then, the long term average distortion under the FFP for any general i.i.d. energy arrivals, $d(\tilde{\bm g})$, satisfies
\begin{align}
d(\tilde{\bm g})\leq\lim_{n\rightarrow\infty}\hat{\mathcal{D}}_n(\tilde{\bm g})
\end{align}
\end{lemma}
Using (\ref{eq_dist_lower_bound}), (\ref{eq_ub_dist}), and Lemma~\ref{thm_bern_iid}, we have
\begin{align}
f(\mu)\leq d^*\leq d(\tilde{\bm g})\leq f(\mu)+\frac{\sigma_s^2}{2}
\end{align}

\section{Sampling Costs: Proof of Theorem~\ref{thm_proc}}

\subsection{Lower Bounding $d_\epsilon^*$}

In this section, we derive the lower bound in (\ref{eq_dist_proc_lower_bound}). Following the same lines as in Section~\ref{sec_lb_dist}, we remove the battery limit and consider the offline setting. We also apply the change of variables: $\bar{g}_t\triangleq\theta_t g_t~\forall t$. The feasible set $\mathcal{F}^\epsilon$ now becomes
\begin{align}
\sum_{t=1}^n\theta_t\epsilon+\bar{g}_t\leq \sum_{t=1}^nE_t,\quad\forall n;\quad 0\leq\theta_t\leq1,\quad\forall t
\end{align}
Next, we use the strong law of large numbers and apply the same $(\delta,n)$ argument as in Section~\ref{sec_lb_dist}. Fixing a $(\delta,n)$ pair, the objective function is given by
\begin{align} \label{eq_dist_h_proc}
\frac{1}{n}\sum_{t=1}^n(1-\theta_t)\sigma_s^2+\frac{\theta_t\sigma_s^2}{1+\frac{\bar{g}_t}{\theta_t\sigma_c^2}}\triangleq\frac{1}{n}\sum_{t=1}^nh\left(\theta_t,\bar{g}_t\right)
\end{align}
It is direct to see that $h$ is jointly convex in $(\theta_t,\bar{g}_t)$ since the second added term is the perspective of the convex function $f(\bar{g}_t)$ \cite{boyd}. Therefore, the optimal power allocation minimizing the objective function is $\theta_t\epsilon+\bar{g}_t=\mu+\delta,~1\leq t\leq n$ \cite{boyd} (see also \cite{jingP2P}). We denote this optimal offline solution by $f_\epsilon(\mu+\delta)$ as defined in (\ref{eq_dist_proc_lower_bound}). We then have $d_\epsilon^*\geq f_\epsilon(\mu+\delta)$; we take $\delta$ down to 0 by taking $n$ infinitely large. Therefore, (\ref{eq_dist_proc_lower_bound}) holds.

\subsection{Upper Bounding $d_\epsilon^*$: Bernoulli Energy Arrivals}

In this section, we derive an upper bound on $d_\epsilon^*$. Following the same steps as in Section~\ref{sec_ub_dist}, we first consider Bernoulli energy arrivals as in (\ref{eq_bern_energy}). In this case we have
\begin{align} \label{eq_renewal_proc}
\lim_{n\rightarrow\infty}\hat{\mathcal{D}}_n^\epsilon\left({\bm \theta},{\bm g}\right) =\frac{1}{\mathbb{E}[L]}\mathbb{E}\left[\sum_{t=1}^L(1-\theta_t)\sigma_s^2+\frac{\theta_t\sigma_s^2}{1+g_t/\sigma_c^2}\right]~\text{a.s.}
\end{align}
where $\hat{\mathcal{D}}_n^\epsilon\left({\bm \theta},{\bm g}\right)$ is the $n$-horizon average distortion under Bernoulli arrivals. Next, we upper bound the long term average distortion in this case by substituting the FFP in (\ref{eq_ffp_dist_proc}) setting
\begin{align} \label{eq_ffp_bern_proc}
\tilde{\theta}_t(\epsilon+\tilde{g}_t)&=p(1-p)^{t-1}B =(1-p)^{t-1}\mu
\end{align}
for all time slots $t$. Note that the average minimal distortion in time slot $t$ is given by $f_\epsilon\left((1-p)^{t-1}\mu\right)$. We have the following lemma regarding $f_\epsilon$. The proof follows by convexity of $h$ and is omitted for space limits.

\begin{lemma} \label{thm_dist_F_proc_cvx}
The function $f_\epsilon$ is convex and non-increasing.
\end{lemma}


Next, following the same steps used in showing (\ref{eq_ub_dist}), by (\ref{eq_renewal_proc}) and (\ref{eq_ffp_bern_proc}), we have
\begin{align} \label{eq_ub_dist_proc}
\lim_{n\rightarrow\infty}\hat{\mathcal{D}}_n^\epsilon\left(\tilde{\bm \theta},\tilde{\bm g}\right) \leq f_\epsilon(\mu)+\frac{\sigma_s^2}{2} 
\end{align}
where step $(a)$ in (\ref{eq_ub_dist}) follows by  Lemma~\ref{thm_dist_F_proc_cvx}. Finally, we use the above result to bound the distortion for general i.i.d. arrivals under the FFP. We basically extend the statement of Lemma~\ref{thm_bern_iid} to the case with sampling costs since $f_\epsilon$ is convex and monotone. We then have
\begin{align} \label{eq_bern_iid_proc}
d_\epsilon\left(\tilde{\bm \theta},\tilde{\bm g}\right)\leq\lim_{n\rightarrow\infty}\hat{\mathcal{D}}_n^\epsilon\left(\tilde{\bm \theta},\tilde{\bm g}\right)
\end{align}
Using (\ref{eq_dist_proc_lower_bound}), (\ref{eq_ub_dist_proc}), and (\ref{eq_bern_iid_proc}), we have
\begin{align}
f_\epsilon(\mu)\leq d_\epsilon^*\leq d_\epsilon\left(\tilde{\bm \theta},\tilde{\bm g}\right)\leq f_\epsilon(\mu)+\frac{\sigma_s^2}{2}
\end{align}

It now remains to show that the FFP corresponds to (\ref{eq_ffp_dist_proc_th}) and (\ref{eq_ffp_dist_proc_pwr}). Towards that end, we solve $f_\epsilon\left(qb_t\right)$ for $\theta$ and $\bar{g}$. We first make the substitution $\bar{g}=qb_t-\theta\epsilon$ into the objective function. The problem now becomes
\begin{align}
\min_{0\leq\theta\leq\min\left\{1,qb_t/\epsilon\right\}}\quad\frac{\theta}{1-\frac{\epsilon}{\sigma_c^2}+\frac{qb_t}{\theta\sigma_c^2}}-\theta
\end{align}
where the constraint $\theta\leq qb_t/\epsilon$ ensures non-negativity of $\bar{g}$. One can show that the objective function above is convex in $\theta$. Hence, we take the derivative, equate to 0, solve for $\theta$, and then project the solution onto the feasible set to get the optimal solution of this problem \cite{boyd}. This gives (\ref{eq_ffp_dist_proc_th}); while (\ref{eq_ffp_dist_proc_pwr}) is directly derived by substituting $g=\frac{qb_t}{\theta}-\epsilon$.

\begin{figure}[t]
\center\includegraphics[scale=.45]{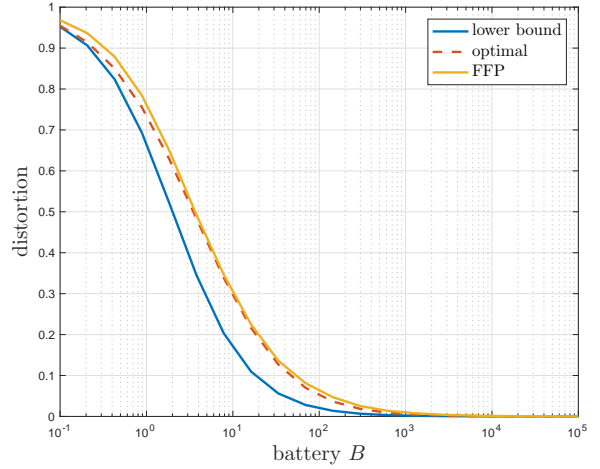}
\caption{Performance of the FFP with no sampling costs.}
\label{fig_ex_dist}
\vspace{-.15in}
\end{figure}

\section{Numerical Results}

In this section we present some examples to illustrate our results. We set both $\sigma_s^2$ and $\sigma_c^2$ to unity, and consider a system with Bernoulli energy arrivals with probability $p=0.5$. In Fig.~\ref{fig_ex_dist}, we plot the lower bound on the long term average distortion for the problem without sampling costs along with the FFP, against the battery size $B$. We also plot the optimal solution in this scenario. We see that the FFP performs very close to the optimal policy. We note that the empirical gap between the optimal policy and the FFP is no larger than $0.03$, while the empirical gap between the lower bound and the FFP is no larger than $0.15$, which is lower than the theoretical gap of $0.5$ in Theorem~\ref{thm_dist}.

\begin{figure}[t]
\center\includegraphics[scale=.45]{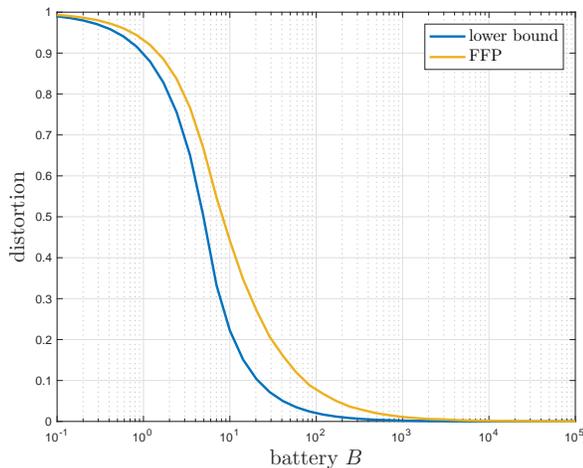}
\caption{Performance of the FFP with sampling costs.}
\label{fig_ex_proc}
\vspace{-.15in}
\end{figure}

In Fig.~\ref{fig_ex_proc}, we plot the same curves for the problem with sampling costs. We set the sampling cost $\epsilon=1.5$. We notice that the distortion levels are higher in general when compared to the case without sampling costs, which is mainly due to having some energy spent in sampling instead of reducing distortion. The empirical gap in this case is $0.22$, which is lower than the theoretical gap of $0.5$ in Theorem~\ref{thm_proc}.


\bibliographystyle{unsrt}

\end{document}